\documentclass[letterpaper,11pt]{article}
\usepackage[hmargin=1.1in,vmargin=1.1in]{geometry}
\usepackage{amsmath}
\usepackage{graphicx}
\usepackage{bm}
\usepackage{epsfig}
\usepackage{amsmath}
\usepackage{amssymb}
\usepackage{xcolor}
\colorlet{linkequation}{blue}
\usepackage[colorlinks]{hyperref}
\newcommand*{\refeq}[1]{%
	\begingroup
	\hypersetup{
		linkcolor=linkequation,
		linkbordercolor=linkequation,
	}%
	\ref{#1}%
	\endgroup
} 
\newcommand{\be}{\begin{equation}}
	\newcommand{\ee}{\end{equation}}
\newcommand{\bea}{\begin{eqnarray}}
	\newcommand{\eea}{\end{eqnarray}}

\title{Trans-Planckian Effect in $f(R)$ Cosmology}
\author{S. Cheraghchi$^{1,2}$, F. Shojai\footnote{Corresponding author: fshojai@ut.ac.ir} $^{2}$, M.H. Abbassi$^3$\\
$^1$Faculty of Mathematics and Computer Science, Transilvania University, \\ Iuliu Maniu Str. 50,
500091 Brasov, Romania\\
$^2$Department of Physics, University of Tehran,\\ P.O. Box 14395-547, Tehran, Iran.\\
$^3$Department of Physics, School of Science, Tarbiat Modares University,\\ P.O. Box 14155-4838, Tehran, Iran\\	}
\begin{document}
\maketitle
\begin{abstract}
Apart from the assumption that the inflation started at an infinite time in the past, the more realistic initial state of the quantum fluctuations  is described by a  mixed quantum state imposed at a finite  value of the initial time.
One of the most important non-trivial vacua is the $\alpha$-vacuum, which is specified by a momentum cutoff $\Lambda$ \cite{Danielsson:2002kx}.  As a consequence,  
the initial condition is imposed at different initial times for the different $k$-modes. This modifies the amplitude of the quantum fluctuations, and thus the corresponding power spectra. 
In this paper, we consider the imprint of the $\alpha$-vacuum state on the power spectrum of scalar perturbations in a generic $f(R)$ gravity by assuming an ultraviolet cutoff $\Lambda$. As a specific model, we consider the Starobinsky model and find the trans-Planckian power spectrum.  We find that the leading order corrections to the scalar power spectra  in $f(R)$ gravity have an oscillatory behavior  as in general relativity \cite{Lim}, and furthermore, the results are in sufficient agreement with the $\Lambda$CDM model.\\
\end{abstract}
\section{Introduction}
The inflationary paradigm not only solves the conceptual problems of the standard Big Bang cosmology but also provides an explanation for the origin of the Large Scale Structures (LSSs) of the Universe \cite{Mukhanov:1981xt}. In the other words, the seeds of the galaxies and galaxy clusters we observe today are the fluctuations of the inflaton field that are generated during inflation. This mechanism is based on the fact that the fluctuation modes, which are well inside the Hubble radius at early times, become larger than the Hubble radius at the end of inflation. Eventually, these fluctuations leave traces in the cosmic microwave background radiation (CMB) that can be observed today.\\
However, if inflation lasts only slightly more than 70 $e$-fold \footnote{For example, in models where inflation starts near the scale of grand unification.}, then the wavelengths of the fluctuations that are currently inside the Hubble radius could have been smaller than the Planck scale at the onset of inflation \cite{Brandenberger:2012aj}. Since the physics on scales smaller than the Planck scale is still unknown, new physics is needed to understand the origin and evolution of the fluctuation modes at very early times. This problem is called the trans-Planckian problem of Inflationary Cosmology \cite{Martin:2000xs}, which suggests that the initial conditions for inflation can be placed on trans-Planckian scales, so that there would be a signature of these scales in the cosmological measurements.\\
The Trans-Planckian Censorship Conjecture (TCC) \cite{TCC} states that the trans-Planckian scales must be hidden by a Hubble horizon. According to the TCC, certain types of cosmological models may be inconsistent during inflation unless certain conditions are met.
The TCC is able to impose bounds on the number of e-folds and on the energy density during the inflation \cite{Bedroya}. 
In terms of these bounds, if the primordial gravitational waves are detected, most inflationary models would be ruled out, unless we make some modifications to this scenario, such as imposing an initial time for the inflationary era. The TCC also prevents us from doing the same study of the fluctuation modes in two scales, below and beyond Planck's scales. Therefore, it would be reasonable to assume a pre-inflationary era and an initial time for the onset of inflation \cite{Suddhasattwa Brahma}.
The choice of an $\alpha$-vacuum initial condition provides a mechanism to avoid the danger of rejecting the inflationary scenario. Thus, we can find some inflationary models that are still compatible with the TCC.\\
There are several approaches to tackle the trans-Planckian problem of inflationary cosmology, see \cite{Danielsson:2002kx, Martin:2000xs, Brandenberger:2000wr,Niemeyer:2000eh,Kaloper:2002uj,Burgess:2002ub,Burgess:2003zw, 2020} for original related works. In \cite{Danielsson:2002kx}, which is followed in the present work, the trans-Planckian problem was discussed from the point of view of the vacuum choice.
Traditionally, the initial conditions for inflaton fluctuations are chosen to be the so-called Bunch-Davies (BD) vacuum \cite{Bunch:1978yq}. In the BD vacuum prescription, it is implicitly assumed that the infinite past, in which  the space-time resembles Minkowskian,  is accessible. The main idea of the approach introduced in \cite{Danielsson:2002kx} is that since the duration of inflation is finite, therefore, imposing initial conditions in the infinite past may lead to inaccurate results. Furthermore, one cannot follow a given fluctuation mode to infinitely small scales, because the currently known physics is limited to scales larger than the Planck scale. Remarkably, the trans-Planckian effects could appear in the CMB spectrum if some non-standard initial conditions are chosen.\\
The inaccessibility of the Minkowskian vacuum in the infinite past is avoided by introducing a momentum cutoff $\Lambda$ 
in the description of Danielsson's $\alpha$-vacua \cite{Danielsson:2002kx}. As mentioned above, the evolution of the fluctuation modes is assumed to begin when the physical momentum $k$ associated with a given fluctuation mode satisfies
$k=a(t_i)\Lambda$, where $a(t_i)$ is the scale factor at the initial time, $t_i$, when the initial condition is imposed. 
Considering a general $f(R)$ gravity, we are interested here in investigating the effect of $\alpha$-vacua in  slow roll inflation.\\
Among the proposed $f(R)$ gravity models, there are several examples that can provide a good description of the inflationary era. One of the relatively old models of modified gravity is the Starobinsky model, $f(R)=R+\mu_0 R^2$, which can successfully describe the early inflationary era without any additional scalar field \cite{Starobinsky}. Another example is the power law gravity $f(R)$, which can unify the early and late time acceleration phases of the universe \cite{power law 1},\cite{power law 2}. The exponential $f(R)$ gravity as a viable modified gravity model has been proposed in \cite{exp model 1}, \cite{exp model 2}, \cite{exp model 3} and then extensively considered in several articles. The main feature of this model is that it has only one parameter more than the $\Lambda$CDM model and passes the viability conditions such as stability, ghost-free and the presence of the matter-dominated stage (see \cite{exp model 4} and references therein).\\
This paper is organized as follows: In Section 2, we discuss the problem of the initial vacuum in the
$f(R)$ inflationary cosmology. Then, in Section 3, we will consider the primordial power spectrum of cosmological scalar fluctuations with the $\alpha$-vacuum initial condition. In Section 4, we focus on a specific model of modified gravity, the Starobinsky  model, and derive the corresponding power spectrum. Section 5 is devoted to the comparison of our derived power spectrum with the observational data. Finally, in Section 6, we provide some discussion of our results and further directions that can be explored in the future works, are mentioned.
\section{Vacuum ambiguity in $f(R)$ inflationary cosmology}\label{sec1}
We consider one of the most general modifications of general relativity, $f(R)$ gravity, where $f$ is a  smooth function of the Ricci scalar, $R$. The action of this theory is given by
\begin{align}\label{action1}
	S=\frac{1}{2}\int{\sqrt{-g}f(R)d^4x}+S_m(g_{\mu\nu},\psi)
\end{align}
where $\kappa^2=8\pi G$ is set to unity. $S_m$ is the matter action that depends on the matter field $\psi$ which  is minimally coupled to gravity. In the following, we consider the spatially flat \\
Friedmann–Lemaitre–Robertson–Walker (FLRW) space-time with the scale factor $a(t)$ described by $ds^{2}=-dt^{2}+a(t)^{2}d\vec{x}^{2}$.\\ 
To study the effects of a non-trivial initial vacuum state on the power spectrum of scalar perturbations, 
we will expand the above action up to the second order in the curvature perturbations. Before doing so, note that due to the extra degree of freedom in $f(R)$ gravity, there are more gauge invariant quantities than those are defined in general relativity. For example, regarding the scalar curvature perturbations, with $F(R)\equiv\partial f/\partial R$, one can define the following gauge invariant quantities \cite{living review}
\begin{align}\label{curv}
\begin{split}
\mathcal{R}=\Psi+\frac{H}{P+\rho}\delta q\hspace{0.5in}
\mathcal{R}_{\delta F}=\Psi-\frac{H}{\dot{F}}\delta F
\end{split}
\end{align}
where $\rho$, $P$  and $H$ are the energy density, pressure and Hubble parameter respectively and dot denotes the time derivative. 
Also, $\delta q=(\rho+P)v$ in which $v$ is the velocity
potential of matter.  $\Psi$ represents the trace part of the spatial scalar perturbation of the metric and $\mathcal{R}$ is the standard curvature perturbation defined in general relativity. $\mathcal{R}_{\delta F}$ appears due to the arbitrary choice of the $f$ function.
The evolution equation governing for each of the scalar curvature perturbations (\ref{curv}) is known as the Mukhanov–Sasaki equation \cite{Hwang}, which is usually derived by some gauge fixing conditions\footnote{Note that the Mukhanov-Sasaki equation is a gauge invariant equation, i.e. a change of the gauge does not have any effect on the equation. However, the calculations to obtain the equation are greatly simplified by choosing an appropriate gauge. 
In $f(R)$ gravity, choosing the gauge condition $\delta F=0$, as a specific choice, effectively removes some degrees of freedom and the Mukhanov-Sasaki equation is easily obtained. Even for a general $f(R,\phi)$ gravity, one can introduce a new gauge by which the generalized Mukhanov–Sasaki equation is derived straightforwardly \cite{our 2nd paper}.}
To quantize the fluctuations, it is necessary to obtain the quadratic action for the curvature perturbation $\mathcal{R}$. Expanding the action (\ref{action1}) up to the second order in the perturbation, we get \footnote{Here we do not specify the gauge. Therefore, $\mathcal{R}$ can be any of the scalar curvature perturbations introduced in (\ref{curv}).} \cite{living review}
\begin{align}\label{expansion of the action1}
	 S^{(2)}=\int{d^4x a^3 Q_s\bigg(\frac{1}{2}\dot{\mathcal{R}}^2-\frac{1}{2 a^2}(\nabla\mathcal{R})^2\bigg)}
\end{align}
where 
\begin{align}\label{Qs}
Q_s\equiv\frac{3\dot F^2}{2F\left( H+\frac{\dot F}{2F}\right)^2}
\end{align}
By defining $z=a\sqrt{Q_s}$  and the Mukhanov- Sasaki variable by $v=z\mathcal{R}$, the action (\ref{Qs}) can be simplified as
\begin{align}\label{changed exapnsion of action}
S^{(2)}=\int{d\tau d^3x\bigg(\frac{1}{2}v'^2-\frac{1}{2}(\nabla v)^2+\frac{1}{2}\frac{z''}{z}v^2\bigg)}
\end{align}
where prime denotes the time derivative with respect to the conformal time given by  $d\tau=dt/a$ and the conjugate momentum  corresponding to the field $v$ is $\pi = v'$.
Variation of the above action with respect to $v(\tau,\vec{x})$ gives the Mukhanov-Sasaki equation. Then using a Fourier decomposition, the mode function $v_{\vec{k}}$ depends only on the magnitude of the wavenumber $\vec{k}$ and satisfies the following equation
\begin{eqnarray}\label{SectionTwoV}
v_{k}''+(k^2-\frac{z''}{z})v_{k}=0
\end{eqnarray}
which is the equation of a harmonic oscillator with a time-dependent frequency. 
To quantize the theory, we convert the Fourier modes $v_{\vec{k}}$ and their conjugate momentum $\pi_{\vec{k}}$ into  quantum operators $\hat{v}_{\vec{k}}$ and $\hat{\pi}_{\vec{k}}$ that satisfy the following standard equal-time commutation relations 
\begin{eqnarray}\label{SectionTwoVIII}
\begin{split}
&[\hat{v}_{\vec{k}}(\tau),\hat{\pi}_{\vec{k'}}(\tau)]=i\delta^{(3)}(\vec{k}+\vec{k'})\\&
[\hat{v}_{\vec{k}}(\tau),\hat{v}_{\vec{k'}}(\tau)]=[\hat{\pi}_{\vec{k}}(\tau),\hat{\pi}_{\vec{k'}}(\tau)]=0
\end{split}
\end{eqnarray}
The above operators can be written as
\begin{eqnarray}\label{SectionTwoIX}
\begin{split}
&\hat{v}_{\vec{k}}(\tau)=\frac{1}{\sqrt{2k}}\left[\hat{a}_{\vec{k}}(\tau)+\hat{a}^{\dagger}_{-\vec{k}}(\tau)\right]\\&
\hat{\pi}_{\vec{k}}(\tau)=-i\sqrt{\frac{k}{2}}\left[\hat{a}_{\vec{k}}(\tau)-\hat{a}^{\dagger}_{-\vec{k}}(\tau)\right]
\end{split}
\end{eqnarray}
in which the time-dependent operators $\hat{a}^{\dagger}_{\vec{k}}$ and  $\hat{a}_{\vec{k}}$ can be interpreted as creation and annihilation operators satisfying  
\begin{eqnarray}\label{SectionTwoXI}
\begin{split}
&[\hat{a}_{\vec{k}}(\tau),\hat{a}^{\dagger}_{\vec{k'}}(\tau)]=\delta^{(3)}(\vec{k}-\vec{k'})\\&
[\hat{a}_{\vec{k}}(\tau),\hat{a}_{\vec{k'}}(\tau)]=[\hat{a}^{\dagger}_{\vec{k}}(\tau),\hat{a}^{\dagger}_{\vec{k'}}(\tau)]=0
\end{split}
\end{eqnarray}
by (\ref{SectionTwoVIII}). This leads to a normalization relation which will be considered below.
The time evolution of these operators is described in the Heisenberg picture and can be written in the most general form as 
\begin{eqnarray}\label{SectionTwoXIII}
\begin{split}
&\hat{a}_{\vec{k}}(\tau)=\alpha_k(\tau)\hat{a}_{\vec{k}}(\tau_i)+\beta_k(\tau)\hat{a}^{\dagger}_{-\vec{k}}(\tau_i)\\&
\hat{a}^{\dagger}_{-\vec{k}}(\tau)=\alpha^{\star}_k(\tau)\hat{a}^{\dagger}_{-\vec{k}}(\tau_i)+\beta^{\star}_k(\tau)\hat{a}_{\vec{k}}(\tau_i)
\end{split}
\end{eqnarray}
These are the Bogoliubov transformations, which describe the mixing of creation and annihilation operators with time and $\tau_i$ as an arbitrary initial time. The functions $\alpha_k(\tau)$ and $\beta_k(\tau)$ are the Bogoliubov coefficients and the asterisk denotes complex conjugation. Since the creation and annihilation operators obey the commutation relations (\ref{SectionTwoXI}),
the Bogoliubov coefficients satisfy the following normalization relation
\begin{eqnarray}\label{SectionTwoXIV}
|{\alpha_k(\tau)}^2|-|{\beta_k(\tau)}|^2=1
\end{eqnarray}
Corresponding to the field operator $\hat{v}_{\vec{k}}(\tau)$ and its canonical momentum $\hat{\pi}_{\vec{k}}(\tau)$, the mode functions $f_k(\tau)$ and $g_k(\tau)$ can be defined by the relations
\begin{eqnarray}\label{SectionTwoXV}
\begin{split}
&\hat{v}_{\vec{k}}(\tau)=f_k(\tau)\hat{a}_{\vec{k}}(\tau_i)+f^{\star}_k(\tau)\hat{a}^{\dagger}_{-\vec{k}}(\tau_i)\\&
\hat{\pi}_{\vec{k}}(\tau)=-i\left[g_k(\tau)\hat{a}_{\vec{k}}(\tau_i)-g^{\star}_k(\tau)\hat{a}^{\dagger}_{-\vec{k}}(\tau_i)\right]
\end{split}
\end{eqnarray}
where $f_k(\tau)$ should satisfy the mode function equation (\ref{SectionTwoV}). Substituting the expansion (\ref{SectionTwoXIII}) into (\ref{SectionTwoIX}) and then comparing with (\ref{SectionTwoXV}), we obtain
\begin{eqnarray}\label{SectionTwoXVI}
\begin{split}
&f_k(\tau)=\frac{1}{\sqrt{2k}}\left[(\alpha_k(\tau)+\beta^{\star}_k(\tau)\right]\\&
g_k(\tau)=\sqrt{\frac{k}{2}}\left[\alpha_k(\tau)-\beta^{\star}_k(\tau)\right]=if'_k(\tau)
\end{split}
\end{eqnarray}
The commutation relations (\ref{SectionTwoVIII}) and (\ref{SectionTwoXI}) also give the following Wronskian condition for the mode functions
\begin{eqnarray}\label{SectionTwoXVII}
g_k(\tau)f^{\star}_k(\tau)+g^{\star}_k(\tau)f_k(\tau)=i\left[f'_k(\tau)f^{\star}_k(\tau)-f'^{\star}_k(\tau)f_k(\tau)\right]=1
\end{eqnarray}
To completely fix the mode functions, we should impose another condition in addition to (\ref{SectionTwoXVII}).
This is an initial condition and comes from the vacuum state of quantum fluctuations.
The natural choice of the initial vacuum is given by
\begin{eqnarray}\label{SectionTwoXVIII}
\hat{a}_{\vec{k}}(\tau_i)|0,\tau_i\rangle=0
\end{eqnarray}
By inserting the annihilation operator from (\ref{SectionTwoXIII}), the above relation becomes
\begin{eqnarray}\label{SectionTwoXIX}
\alpha_k(\tau_i)\hat{a}_{\vec{k}}(\tau_i)|0,\tau_i\rangle+\beta_k(\tau_i)\hat{a}^{\dagger}_{-\vec{k}}(\tau_i)|0,\tau_i\rangle=0
\end{eqnarray}
Considering (\ref{SectionTwoXVIII}), the first term on the left hand side of equation (\ref{SectionTwoXIX}) would be zero. Therefore, to have a consistent definition for the vacuum state, we require  that $\beta_k(\tau_i)=0$. As $\tau_i\rightarrow \infty$, all perturbations are deep inside the Hubble horizon and therefore, the frequency of the harmonic oscillator  in (\ref{SectionTwoV}) would be time independent. The corresponding vacuum state is called BD vacuum which is the standard vacuum state. For $\alpha$-vacua, the initial state is imposed at some arbitrary finite time $\tau_i$ and it is in fact an excited state obtained by applying the Bogoliubov transformation to the BD vacuum.  In this way,  there is a pre-inflationary period, after which the inflationary expansion begins.
Note that the $\alpha$-vacua prescriptions differ significantly depending on the choice of $\tau_i$ and also of the Bogoliubov coefficients in (\ref{SectionTwoXIX}). In the simplest case, the condition (\ref{SectionTwoXVIII}) is satisfied at $\tau_i$ and thus $\beta_k(\tau_i)=0$. According to the proposal of Danielsson \cite{Danielsson:2002kx}, the initial time  $\tau_i$ can be defined by introducing a physical momentum cutoff $\Lambda$ such that the evolution of the perturbation modes starts as soon as the perturbation wavenumber satisfies
$k= a(\tau_i )\Lambda$. We will return to this point in the next section.
\section{Quantum fluctuations in $f(R)$ slow-roll inflation}
To study the evolution of the mode functions during slow-roll inflation, we introduce the corresponding slow-roll parameters  \cite{Hwang}:
\begin{align}\label{slow roll parameters}
\epsilon_1\equiv-\frac{\dot{H}}{H^2},\hspace{0.3in}\epsilon_3\equiv\frac{\dot{F}}{2FH},\hspace{0.3in}{\epsilon_4\equiv\frac{\ddot{F}}{H\dot{F}}}
\end{align}
According to (\ref{SectionTwoV}), we must first find $z''/z$ in terms of the slow-roll parameters. To do this, we rewrite it as follows
\begin{align}
\begin{split}
\frac{z''}{z}=\frac{1}{H^2\tau^2(1-\epsilon_1)^2}\left(H\frac{\dot{z}}{z}+\frac{\ddot{z}}{z}\right)
\end{split}
\end{align}
for which we have used 
\begin{align}\label{hubbel parameter}
\mathcal{H}=-\frac{1}{(1-\epsilon_1)\tau}
\end{align}
where $\mathcal{H}=a'/a=a H$ and the above relation is obtained by integrating the first relation in (\ref{slow roll parameters}) and assuming that  $\epsilon_1$ is nearly constant. We also assume here that all slow-roll parameters defined by (\ref{slow roll parameters}) are approximately constant. Thus,
 up to the leading order  in the slow-roll parameters, $z''/z$ reduces to  \cite{living review}
\begin{align}
\begin{split}
\frac{z''}{z}=\frac{1}{\tau^2(1-\epsilon_1)^2}\bigg(2 +2 \epsilon_1-3 \epsilon_3+3 \epsilon_4-\epsilon_1 \epsilon_3+\epsilon_1\epsilon_4+\epsilon_3^2-2 \epsilon_3\epsilon_4+\epsilon_4^2\bigg)
\end{split}
\end{align}
where as mentioned before, $z=a\sqrt{Q_s}$  and $Q_s$ is defined in (\ref{Qs}).
It is usually more convenient to express $z''/z$ in the following form  
\begin{align}\label{zz}
\frac{z''}{z}=\frac{\nu_\mathcal{R}^2-\frac{1}{4}}{\tau^2}\hspace{0.5in}\nu_\mathcal{R}^2=\frac{1}{4}+\frac{(1+\epsilon_1-\epsilon_3+\epsilon_4)(2-\epsilon_3+\epsilon_4)}{(1-\epsilon_1)^2}
\end{align}
Up to the first order in slow roll parameters: $\nu_\mathcal{R}\sim 3/2+2\epsilon_1-\epsilon_3+\epsilon_4$. Therefore, the general solution of (\ref{SectionTwoV}) can be expressed as a linear combination of the first and second kind Hankel functions, $H_{\nu_\mathcal{R}}^{(1)}$ and $H_{\nu_\mathcal{R}}^{(2)}$
\begin{align}\label{sol eom u}
v_k=\frac{\sqrt{\pi |\tau|}}{2} e^{i\frac{\pi}{4}(1+2\nu_\mathcal{R})}\bigg(C_1 H_{\nu_\mathcal{R}}^{(1)}(k|\tau|)+C_2 H_{\nu_\mathcal{R}}^{(2)}(k|\tau|)\bigg)
\end{align} 
where $C_1$ and $C_2$ are $k$-dependent integration constants. This gives the mode functions $f_k(\tau)$ and $g_k(\tau)$, so the quantization procedure is straightforward. Substituting (\ref{sol eom u}) in (\ref{SectionTwoXV})-(\ref{SectionTwoXVI}) yields the following mode functions
\begin{eqnarray}
\begin{split}
&f_k(\tau)=v_k(\tau)\\
&g_k(\tau)=if'_k(\tau)=\\
&\frac{i}{\sqrt{|\tau|}}e^{i\frac{\pi}{4}(1+2\nu_\mathcal{R})}\bigg(2C_1 k\tau H_{\nu_\mathcal{R}-1}^{(1)}(|k\tau|)+C_1(1-2\nu_\mathcal{R})H_{\nu_\mathcal{R}}^{(1)}(|k\tau|)+\\
&2C_2 k\tau H_{\nu_\mathcal{R}-1}^{(2)}(|k\tau|)+C_2(1-2\nu_\mathcal{R})H_{\nu_\mathcal{R}}^{(2)}(|k\tau|)\bigg)
\end{split}
\end{eqnarray}
Now, using (\ref{SectionTwoXVI}), by a simple calculation one can find the Bogoliubov coefficients as
\begin{eqnarray}\label{Bogo co}
\begin{split}
&\alpha_k(\tau)=\frac{1}{2}\big(\sqrt{2k}f_k(\tau)+\sqrt{\frac{2}{k}}g_k(\tau)\big)=\\
&e^{i\frac{\pi}{4}(1+2\nu_\mathcal{R})}\frac{1}{4}\sqrt{\frac{\pi}{2|k\tau|}}\bigg(2iC_1 k\tau H_{\nu_\mathcal{R}-1}^{(1)}(|k\tau|)-C_1(i-2i\nu_\mathcal{R}+2k\tau)H_{\nu_\mathcal{R}}^{(1)}(|k\tau|)+\\
&2iC_2 k\tau H_{\nu_\mathcal{R}-1}^{(2)}(|k\tau|)-C_2(i-2i\nu_\mathcal{R}+2k\tau)H_{\nu_\mathcal{R}}^{(2)}(|k\tau|)\bigg)\\
&\beta^*_k(\tau)=\frac{1}{2}\big(\sqrt{2k}f_k(\tau)-\sqrt{\frac{2}{k}}g_k(\tau)\big)=\\
&e^{i\frac{\pi}{4}(1+2\nu_\mathcal{R})}\frac{1}{4}\sqrt{\frac{\pi}{2|k\tau|}}\bigg(-2iC_1 k\tau H_{\nu_\mathcal{R}-1}^{(1)}(|k\tau|)+C_1(i-2i\nu_\mathcal{R}-2k\tau)H_{\nu_\mathcal{R}}^{(1)}(|k\tau|)-\\
&2iC_2 k\tau H_{\nu_\mathcal{R}-1}^{(2)}(|k\tau|)+C_2(i-2i\nu_\mathcal{R}-2k\tau)H_{\nu_\mathcal{R}}^{(2)}(|k\tau|)\bigg)
\end{split}
\end{eqnarray}
As mentioned before, the traditional BD vacuum implies that in the infinite past, $k\tau\to-\infty$, the mode functions tend to  $e^{-ik\tau}/\sqrt{2k}$. From this and recalling the asymptotic expression for Hankel functions with large argument 
\begin{align}\label{hh}
H^{(1)}_\nu(x)\to\sqrt{\frac{2}{\pi x}}\exp\left(i x-i\frac{\pi}{2}\nu-i\frac{\pi}{4}\right)\hspace{0.5in}H^{(1)*}_\nu(x)=H^{(2)}_\nu(x)
\end{align}
From (\ref{sol eom u}) we see that $C_1=1$ and $C_2=0$. Thus, the mode function (\ref{sol eom u}) reduces to 
\begin{align}\label{mode function BD}
\textit{f}_k=v_k=\frac{\sqrt{\pi |\tau|}}{2}e^{i\frac{\pi}{4}(1+2\nu_\mathcal{R})} H_{\nu_\mathcal{R}}^{(1)}(|k\tau|)
\end{align}
By applying the initial condition $\beta_k(\tau_i)=0$, the coefficients $C_1$ and $C_2$ are related as follows
\begin{eqnarray}
\begin{split}
&\beta_k^*(\tau_i)\propto -2iC_1 k\tau_i H_{\nu_\mathcal{R}-1}^{(1)}(|k\tau_i|)+C_1(i-2i\nu_\mathcal{R}(\tau_i)-2k\tau_i)H_{\nu_\mathcal{R}}^{(1)}(|k\tau_i|)-\\
&2iC_2 k\tau_i H_{\nu_\mathcal{R}-1}^{(2)}(|k\tau_i|)+C_2(i-2i\nu_\mathcal{R}(\tau_i)-2k\tau_i)H_{\nu_\mathcal{R}}^{(2)}(|k\tau_i|)=0
\end{split}
\end{eqnarray}
Assuming $|k\tau_i|\gg 1$, the asymptotic behavior of the Hankel functions (\ref{hh}) yields
\begin{eqnarray}\label{C2}
\begin{split}
C_2\simeq\frac{iC_1(1-2\nu_\mathcal{R}(\tau_i))e^{-2ik\tau_i}}{4k\tau_i}
\end{split}
\end{eqnarray}
On the other hand, according to the renormalization relation (\ref{SectionTwoXIV}): 
\begin{eqnarray}\label{C1}
\begin{split}
&16(k\tau_i)^2|C_1|^2-8k\tau_i(2\nu_\mathcal{R}(\tau_i)-1)\text{Im}(C_2^* C_1 e^{-2ik\tau_i})+2(1-2\nu_{\mathcal{R}}(\tau_i))^2\text{Re}(C_2^* C_1 e^{-2ik\tau_i})\\
&+(1-2\nu_{\mathcal{R}}(\tau_i))^2(|C_1|^2+|C_2|^2)=16(k\tau_i)^2
\end{split}
\end{eqnarray}
Combining (\ref{C2}) and (\ref{C1}), and given that $|k \tau_i|\gg 1$, the dominant term on the left hand side of (\ref{C1}) is the first term. Thus, it is clear that $|C_1|^2$ is close to unity
\begin{equation}\label{C11}
|C_1|^2\simeq 1.
\end{equation}
For simplicity we set the phase of $C_1$ to $-2k\tau_i$ and then substitute (\ref{C2}) and (\ref{C11}) into (\ref{sol eom u}) to obtain the mode function
\begin{align}
v_k=\frac{\sqrt{\pi |\tau|}}{2} e^{i\frac{\pi}{4}(3+2\nu_\mathcal{R})-2ik\tau_i}\bigg( H_{\nu_\mathcal{R}}^{(1)}(|k\tau|)+\frac{(1-2\nu_{\mathcal{R}})}{4k\tau_i} H_{\nu_\mathcal{R}}^{(2)}(|k\tau|)\bigg)
\end{align} 
It is worth mentioning that considering the modified theory (\ref{action1}), the curvature perturbation remains constant on the super-horizon scales \cite{living review} and as usual, it can be evaluated at the moment of horizon crossing, $k=a H$.\\
To make contact with the observation, we consider the two-point function of the scalar fluctuations produced by inflation. This function is related to the power spectrum of the curvature perturbation by
$\left<\mathcal{R}_{k_1}\mathcal{R}_{k_2}\right>=(2\pi)\delta^3(k_1+k_2)\mathcal{P}_\mathcal{R}(k_1)$
where the power spectrum $\mathcal{P}_\mathcal{R}(k)$ is defined as
\begin{align}\label{power spectrum}
\mathcal{P}_\mathcal{R}(k)\equiv\frac{k^3}{2\pi^2}|\mathcal{R}_k|^2=\frac{k^3}{2\pi^2 z^2}|f_k|^2
\end{align}
Here, we study the power spectrum of scalar fluctuations using the two aforementioned initial vacuum conditions, BD and Danielsson's $\alpha$-vacuum states.
For the BD initial state, using (\ref{hubbel parameter}) and inserting (\ref{mode function BD}) into (\ref{power spectrum}), the power spectrum up to the zero order of the slow roll parameters would be \cite{living review}\begin{align}\label{PS BD Staro}
\mathcal{P}_\mathcal{R}(\tau_k)=\frac{H^2(1-\epsilon_1(\tau_k))^2\Gamma(\nu_\mathcal{R}(\tau_k))^2}{6F\pi^3\epsilon_3^2(\tau_k)}\left(\frac{|k\tau_k|}{2}\right)^{3-2\nu_\mathcal{R}}
\end{align}
where $\tau_k$ is the horizon crossing time of mode $k$ and we have also used $Q_s\simeq 6F \epsilon_3^2$. 
As a first approximation , one can set $\nu_{\mathcal{R}}=3/2$ and use the small argument asymptotic expressions of the  Hankel functions, 
\begin{eqnarray}
H^{(1)}_{\nu}(x)\to -i\Gamma(\nu)\frac{2^{\nu}}{x^{\nu}\pi}\hspace{0.3in}H^{(2)}_{\nu}=H^{(1)*}_{\nu}.
\end{eqnarray}
then the power spectrum of the curvature perturbation will be
\begin{align}
\mathcal{P}_\mathcal{R}(\tau_k)\simeq\frac{1}{6F\epsilon_3^2}\bigg(\frac{H}{2\pi}\bigg)^2
\end{align}
Considering the non-BD initial state, the general mode function (\ref{sol eom u}) can be evaluated at the moment of horizon crossing. It can then be substituted into the power spectrum (\ref{power spectrum}) to obtain
\begin{align}\label{ps}
\mathcal{P}_{\mathcal{R}}^{N-BD}(\tau_k)=\frac{H^2(\tau_k)(1-\epsilon_1(\tau_k))^2\Gamma^2(\nu_\mathcal{R}(\tau_k))}{6F\pi^3\epsilon_3^2(\tau_k)}\bigg(\frac{|k\tau_k|}{2}\bigg)^{3-2\nu_\mathcal{R}(\tau_k)}|C_1-C_2|^2
\end{align}
For the special case of $\alpha$-vacuum, the modification factor up to the first order in slow roll parameters is obtained by comparing (\ref{PS BD Staro}) and (\ref{ps}) as
\begin{eqnarray}
\begin{split}\label{this}
|C_1-C_2|^2=|C_1|^2+|C_2|^2-2\text{Re}(C_1C_2^*)\simeq 1-\frac{(1-2\nu_\mathcal{R}(\tau_i))\sin({2k\tau_i})}{2k\tau_i}
\end{split}
\end{eqnarray}
The power spectrum is simplified by substituting (\ref{this}) in (\ref{ps}):
\begin{align}\label{ps1}
	\begin{split}
	\mathcal{P}^\alpha_\mathcal{R}(\tau_k,k)=\frac{H^2(\tau_k)(1-\epsilon_1(\tau_k))^2\Gamma^2(\nu_\mathcal{R}(\tau_k))}{6F\pi^3\epsilon_3^2(\tau_k)}\bigg(\frac{|k\tau_k|}{2}\bigg)^{3-2\nu_\mathcal{R}(\tau_k)}\bigg[1-\frac{(1-2\nu_\mathcal{R}(\tau_i))\sin({2k\tau_i})}{2k\tau_i}\bigg]
	\end{split}
\end{align}
where the second term in the bracket could be interpreted as 
the correction term of the power spectrum due to the $\alpha$-vacuum state. Therefore, it would be useful to rewrite  equation (\ref{ps1}) in the following form
\begin{align}\label{p}
	\mathcal{P}^\alpha_\mathcal{R}(\tau_k,k)=\mathcal{P}_\mathcal{R}(\tau_k,k)+\delta\mathcal{P}_\mathcal{R}(\tau_k,k)
\end{align}
where $\mathcal{P}_\mathcal{R}(\tau_k,k)$ is the power spectrum (\ref{PS BD Staro}) calculated by the BD initial vacuum state and
\begin{align}\label{dp}
	\begin{split}
	\delta\mathcal{P}_\mathcal{R}(\tau_k,k)\simeq-\frac{H^2(\tau_k)(1-2\nu_\mathcal{R}(\tau_i)-2\epsilon_1(\tau_k))\Gamma^2(\nu_\mathcal{R}(\tau_k))}{6F\pi^3\epsilon_3^2(\tau_k)}\frac{\sin({2k\tau_i})}{2k\tau_i}\bigg(\frac{|k\tau_k|}{2}\bigg)^{3-2\nu_\mathcal{R}(\tau_k)}
	\end{split}
\end{align}
For comparison with the power spectrum (\ref{PS BD Staro}), we expand (\ref{ps1}) up to the first order of the slow roll parameters
\begin{align}
\begin{split}
	\mathcal{P}^\alpha_\mathcal{R}(\tau_k,k=aH)=\frac{H^2}{24\pi^2 F\epsilon_3^2(\tau_k)}\bigg(1+\frac{\sin({2k\tau_i})}{k\tau_i}\bigg)
	\end{split}
\end{align} 
where $\left(|k\tau_k|/2\right)^{3-2\nu_\mathcal{R}}\simeq 1$ for sufficiently late time. As can be seen from this equation, the second term in parentheses shows a small correction due to the initial $\alpha$-vacuum state. This oscillating term introduces a small deviation from the nearly scale-invariant power spectrum and has been found in several models, such as the axion monodromy model \cite{ESilverstein,TKobayashi,LMcAllister} and other models with non-BD initial conditions \cite{Danielsson:2002kx, Brandenberger:2000wr, VBozza, Bouzari nezhad}. In these models, the correction term behaves differently depending on the type of modulation of the primordial power spectrum
\cite{Bouzari nezhad}.
It should be noted that in the correction term (\ref{dp}), the Hubble and slow roll parameters are evaluated at the initial or the horizon crossing time, both of which depend on the fluctuation wavenumber. The correction term amplitude which indicates the power of this modification, is $k$-dependent. The efficiency of this term in the power spectrum of the fluctuations can be controlled by choosing a reasonable value for the pivot scale. Also, by considering the amplitude of the oscillatory term in (\ref{ps1}), one can see that choosing a sufficiently large value of $\tau_i$ or the ultra-violet cutoff $\Lambda$ on the physical momentum prevents naively changes in the power spectrum.
In the next section, we
turn our attention to this point and we see that in the case of modified gravity, the power spectrum of the curvature perturbation depends, as expected, on the model under consideration. Therefore, we must first specify the model in order to find the $k$ dependence of the power spectrum.
In the following,  we will consider the Starobinsky model as a viable model of $f(R)$ gravity. 
\section{Starobinsky's model, $f(R)=R+\mu_0 R^2$}
Let us to consider Starobinsky's model \cite{Starobinsky} in the absence of the scalar field, $f(R)=R+\mu_0 R^2$, where the coefficient $\mu_0$  has the inverse square mass dimension. 
The slow roll parameters of this model are defined as:
\begin{align}\label{s-l relations}
	\begin{split}
&\epsilon_1\equiv-\frac{\dot{H}}{H^2}  \hspace{0.3in} \eta\equiv\frac{\ddot{H}}{2\dot{H}H} \hspace{0.3in}\xi\equiv\frac{\dddot{H}}{H\ddot{H}} \hspace{0.3in}\epsilon_3\equiv\frac{\dot{F}}{2FH}\hspace{0.3in}\epsilon_4\equiv\frac{\ddot{F}}{H\dot F}\\
	\end{split}
\end{align}
where a new definition of $\epsilon_4$ has been introduced. In this inflationary model, $F=1+12\mu_0(2H^2+\dot{H})=1+12\mu_0 H^2(2-\epsilon_1)\sim 24\mu_0 H^2$,  where we have ignored the first-order term because
the curvature is extremely high ($\mu_0 H^2\gg 1$) in the early universe. Substituting $F$ in $\epsilon_3$ defined by (\ref{s-l relations}), we find $\epsilon_3\simeq-\epsilon_1$.  It can also be shown that $\epsilon_4=\dot{H}/H^2+\ddot{H}/H\dot{H}=2\eta-\epsilon_1$\footnote{Note that in the Starobinsky model an additional degree of freedom can be introduced by the definition of $\eta$ as a slow roll parameter. In \cite{living review}, it is shown that under the assumption that $\ddot H\sim 0$, all slow roll parameters are related to $\epsilon_1$, which is proportional to the mass scale, which is the free parameter of the model. In this paper, we do not make this assumption and we define the additional degree of freedom, $\eta$. The degrees of freedom of the model are also increased by the finite initial time of the inflationary era and the $k$-dependence of the Hubble and slow-roll parameters, as we will show in the following.}
Using (\ref{s-l relations}), expanding $Q_s$ (\ref{Qs}) and  ${z''}/{z}$ up to the leading  order of the slow-roll parameters  yields
\begin{align}\label{Qs Staro}
\begin{split}
	&Q_s\simeq 6F \epsilon_1^2\\
	&\frac{z''}{z}=\frac{1/4-\nu_\mathcal{R}^2}{\tau^2}=\mathcal{H}^2(2+2\epsilon_1+6\eta)	
	\end{split}
\end{align}
The last relation is derived in a similar way as (\ref{zz}) and we can easily see that $\nu_\mathcal{R}$ up to the first order of the slow roll parameters is $\nu_\mathcal{R}\simeq 3/2+2\epsilon_1+2\eta$.
Now we assume that Danielsson's $\alpha$-vacuum \cite{Danielsson:2002kx} is used to define the initial conditions of the  fluctuations. As mentioned before, this means that for each mode with wavenumber $k$, there exists a finite $\tau_i$ (or equivalently $t_i$)  such that the  physical momentum of the mode is given by $\Lambda=k/ a(t_i)$ where $\Lambda$ is a constant momentum scale. Therefore, the  initial conditions for each mode depend on its wavenumber. 
 In other words, different fluctuation modes would satisfy the initial condition at different times. To study the $k$-dependence of $\tau_i$, we need to find the $k$-dependence of $H(t_i)$. 
This is done by differentiating the initial condition $\Lambda=k/ a(t_i)$ with respect to $k$ which gives ${dk}/{dt_i}=kH(t_i)$. Combining this result with the definition of the first slow roll parameter given by (\ref{s-l relations}), we find the $k$-dependence of the Hubble parameter up to the leading order of the slow-roll parameters
\begin{align}\label{H}
	\frac{dH}{H}=-\frac{\epsilon_1 dk}{k}
\end{align} 
Assuming that $\epsilon_1$ is very small, we get
\begin{align}
H({t_i})=H_f\big(\frac{k}{k_f}\big)^{-\epsilon_1}
\end{align}
where $H_f$ is the Hubble parameter at the time when the first scale $k_f$ satisfies the initial condition.
Using (\ref{hubbel parameter}), 
we then have 
\begin{align}\label{tauI}
	\tau_i\equiv\tau({t_i})=\frac{-\Lambda}{k H({t_i})(1-\epsilon_1({t_i}))}\simeq\frac{-\Lambda}{k H({t_i})}(1+\epsilon_1({t_i}))
\end{align} 
where we have used the smallness of the slow roll parameter $\epsilon_1({t_i})$. 
Similarly, it would be easy to find the $k$-dependence of the slow roll parameters from the definition of $\eta$ and $\xi$ in (\ref{s-l relations}) and also from equation (\ref{H})
\begin{align}\label{epsilon1i}
\begin{split}	
	&2\eta\frac{dk}{k}=\frac{d(H^2\epsilon_1)}{H^2\epsilon_1}\hspace{0.5in}\xi\frac{dk}{k}=\frac{d(\epsilon_1\eta H^3)}{\epsilon_1\eta H^3}
	\end{split}
\end{align}
Substituting (\ref{H}) into (\ref{epsilon1i}) and then integrating the result gives
\begin{align}
\epsilon_1({t_i})=\epsilon_{1f}\big(\frac{k}{k_f}\big)^{2(\eta+\epsilon_1)}\hspace{0.5in}\eta(t_{i})=\eta_f\big(\frac{k}{k_f}\big)^{\xi-2\eta+\epsilon_1}
\end{align}
where $\epsilon_{1f}$ and $\eta_f$ are the values of $\epsilon_1(t)$ and $\eta(t)$ at the moment when the first scale satisfies the initial condition. On the other hand, we can follow the same procedure as above to find the $k$-dependence of the Hubble and slow roll parameters at the moment of horizon crossing $k=a(t_k)H(t_k)$. Differentiating the horizon crossing condition with respect to $t_k$ gives $dk/dt_k=kH(t_k)(1-\epsilon_1)$. Then using the definition of the first slow roll parameter (\ref{s-l relations}) and its smallness,  we find that $dH/H(t_k)=-\epsilon_1 dk/k$ which gives $H(t_k)=H_l(k/k_l)^{-\epsilon_1}$.
This leads to the $k$-dependence of $\epsilon_1(t_k)$ as
\begin{align}\label{epsilon1k}
2\eta\frac{dk}{k}=\frac{d(H^2(t_k)\epsilon_1(t_k))}{H^2(t_k)\epsilon_1(t_k)}
\end{align}
so that
\begin{equation}
\epsilon_1({t}_k)=\epsilon_{1l}\big(\frac{k}{k_l}\big)^{2(\eta+\epsilon_1)}
\end{equation}
where $\epsilon_{1l}$ is evaluated when the last scale leaves the horizon.
Also, it should be noted that in (\ref{epsilon1i}) and (\ref{epsilon1k}), $\epsilon_1$ and $\eta$ in the exponents are assumed to be constant and will be fitted with data. It is easy to see that $\epsilon_1(t_{i})$ and $\epsilon_1(t_k)$ are related by
\begin{align}\label{epsilonl epsilonf}
\epsilon_{1f}=\epsilon_{1l}\left(\frac{k_f}{k_l}\right)^{2(\epsilon_1+\eta)}
\end{align}
Let us now compute the power spectrum in the Starobinsky model. We take $\alpha$-vacuum as the initial vacuum state.
Substituting  (\ref{Qs Staro}-\ref{epsilon1k}) into (\ref{ps1}), gives the leading order of the power spectrum as
\begin{align}\label{PS}
	\begin{split}
&\mathcal{P}^\alpha_\mathcal{R}=\frac{\Gamma(\nu_\mathcal{R}(\tau_k))^2(1-\epsilon_1(\tau_k))^2}{144\mu_0\pi^3\epsilon_1(\tau_k)^2}\bigg(\frac{|k\tau_k|}{2}\bigg)^{-4(\epsilon_1(\tau_k)+\eta(\tau_k))}\bigg[1+(1+2\eta(\tau_{ik})+2\epsilon_1(\tau_{ik}))\frac{\sin{(2k\tau_i)}}{k\tau_i}\bigg]=\\
&\frac{1}{576\mu_0\pi^2\epsilon_{1l}^2}\bigg(\frac{k}{k_l}\bigg)^{-4(\eta+\epsilon_1)}\bigg\lbrace 1+0.92\epsilon_{1l}\bigg(\frac{k}{k_l}\bigg)^{2(\eta+\epsilon_1)}+2.92\eta_l\bigg(\frac{k}{k_l}\bigg)^{\xi-2\eta+\epsilon_1}+\bigg[1+0.92\epsilon_{1l}\bigg(\frac{k}{k_l}\bigg)^{2(\eta+\epsilon_1)}\\
&+2.92\eta_l\bigg(\frac{k}{k_l}\bigg)^{\xi-2\eta+\epsilon_1}+2\eta_f\bigg(\frac{k}{k_f}\bigg)^{\xi-2\eta+\epsilon_1}+2\epsilon_{1f}\bigg(\frac{k}{k_f}\bigg)^{2(\eta+\epsilon_1)}\bigg]\frac{\sin{\bigg(\frac{2\Lambda}{H_f}\big(\frac{k}{k_f}\big)^{\epsilon_1}\bigg)}}{\frac{\Lambda}{H_f}\big(\frac{k}{k_f}\big)^{\epsilon_1}}\bigg\rbrace
	\end{split}
	\end{align}
where we have used the previously mentioned relations $\epsilon_3\simeq-\epsilon_1$, $\epsilon_4\sim2\eta-\epsilon_1 $
and the following expansion
\begin{align}
\bigg(\frac{|k\tau_k|}{2}\bigg)^{-4(\epsilon_1(\tau_k)+\eta(\tau_k))}(1-\epsilon_1(\tau_k))^2\Gamma(\nu_\mathcal{R}(\tau_k))^2= \frac{\pi}{4}(1+0.92\epsilon_1(\tau_k)+2.92\eta(\tau_k))+\mathcal{O}(\epsilon^2)
\end{align}
in which equation (\ref{hubbel parameter}) has been used to express $|k\tau_k|$ in terms of slow roll parameters.\\
It is now worthwhile to compare the power spectrum of the $\alpha$-vacuum (\refeq{PS}) with that of the BD vacuum in the Starobinsky model.
To make our comparison more clear, we consider only the leading order in the slow-roll approximation
\begin{align}\label{delta Ps}
\frac{\mathcal{P}^\alpha_{\mathcal{R}}(k)}{\mathcal{P}_{\mathcal{R}}(k)}=\frac{\mathcal{P}_{\mathcal{R}}(k)+\delta\mathcal{P}_\mathcal{R}(k)}{\mathcal{P}_{\mathcal{R}}(k)}\simeq 1+\frac{H_f}{\Lambda}\sin{(2\Lambda/H_f)}
\end{align}
This ratio can be compared with the corresponding one in GR obtained in \cite{Bouzari nezhad}  where the authors considered a massless single-field inflationary model with minimal coupling between the inflationary field and gravity. The Starobinsky model and the single scalar field inflationary model can be related by a conformal transformation. 
Thus, the comparison of our results with those of \cite{Bouzari nezhad} may appear to be a comparison of models with different potentials.
The point is that there is still some ambiguity about the physical equivalence of these two frames, the Jordan frame and the Einstein frame \cite{Capozziello}. \\
According to \cite{Bouzari nezhad}
\begin{align}
\frac{\mathcal{P}^{\alpha,GR}_{\mathcal{R}}(k)}{\mathcal{P}_{\mathcal{R}}(k)}\simeq 1+\frac{H_f}{\Lambda}(2\epsilon_{1f}-\eta_f)\sin(2\Lambda/H_f)
\end{align}
in the leading order of the slow roll parameters.
As it is clear form this relation, in contrast to the Starobinsky model, the correction term in the slow roll parameters in GR is of order one.\\
We end up this section by calculating the spectral index of the scalar perturbation 
\begin{align}
	n^{\alpha}_s(k)-1\equiv\frac{k}{\mathcal{P}^\alpha_\mathcal{R}(k)}\frac{d \mathcal{P}^\alpha_\mathcal{R}(k)}{dk}\mid_{k=aH(t_k)}
\end{align}
for the Starobinsky model. Substituting (\ref{PS}) into the above relation, the spectral index up to the leading order in the slow roll parameters would be
\begin{align}\label{nstp}
n^{\alpha}_s(k)-1=-4(\epsilon_1+\eta)+\frac{2\cos{\bigg( 2\frac{\Lambda}{H_f}\bigg)}-\frac{H_f}{\Lambda}\sin{\bigg( 2\frac{\Lambda}{H_f}\bigg)}}{1+\frac{H_f}{\Lambda}\sin{\bigg( 2\frac{\Lambda}{H_f}\bigg)}}\epsilon_1
\end{align}
The corrections due to the N-BD vacuum initial state are shown in the second term on the right hand side of (\refeq{nstp}). It is easy to see that the spectral index is scale independent up to the leading order in the slow roll parameters.
If we compare the above spectral index with the corresponding one resulting from the BD initial state, with respect to \eqref{PS} in the infinite past limit, we can see that $\mathcal{P}^{\alpha}_{\mathcal{R}}\sim\mathcal{P}_\mathcal{R}$.  Thus, it is concluded that in this limit $n_s-1=-4(\epsilon_1+\eta)$\footnote{As mentioned before, by assuming that $\ddot H\sim 0$, we get $\eta=0$ and then $n_s-1=-4\epsilon_1$, which is the same as the result obtained in \cite{living review}.}.
\section{Comparison with data}
\begin{figure}
  \centering
  \includegraphics[width=12cm]{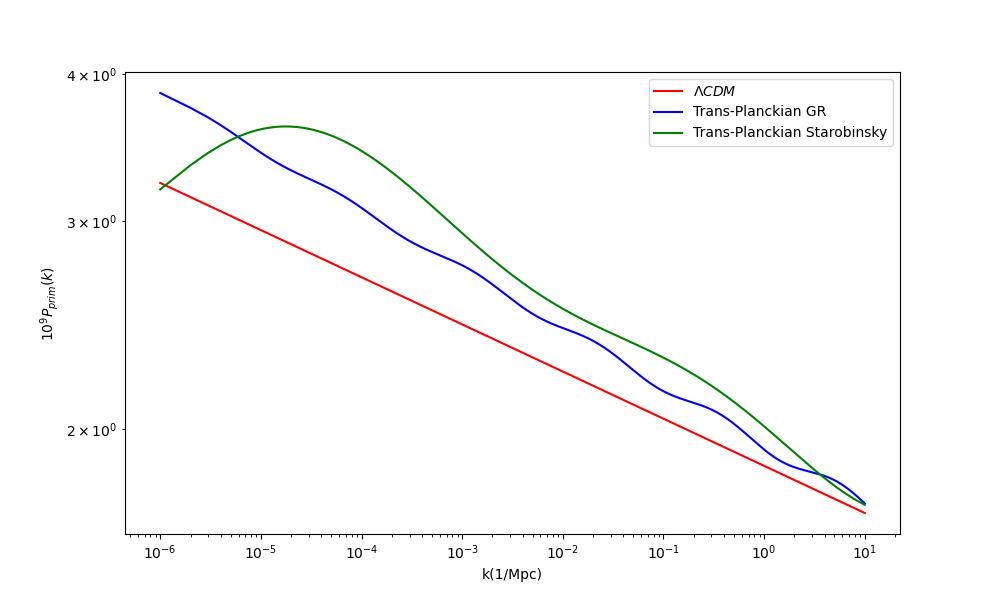}
  \caption{\footnotesize{The red line shows the primordial power spectrum of standard $\Lambda$CDM using the parameters of Planck 2018, the primordial power spectrum of trans-Planckian GR \cite{Bouzari nezhad} is plotted by the blue line for the best-fit parameters found in \cite{Bouzari nezhad} and the green line shows the primordial power spectrum of trans-Planckian Starobisky using the best-fit parameters of Table 1.}}
  \label{fig:prim ps}
\end{figure}

\begin{figure}
  \centering
  \includegraphics[width=12cm]{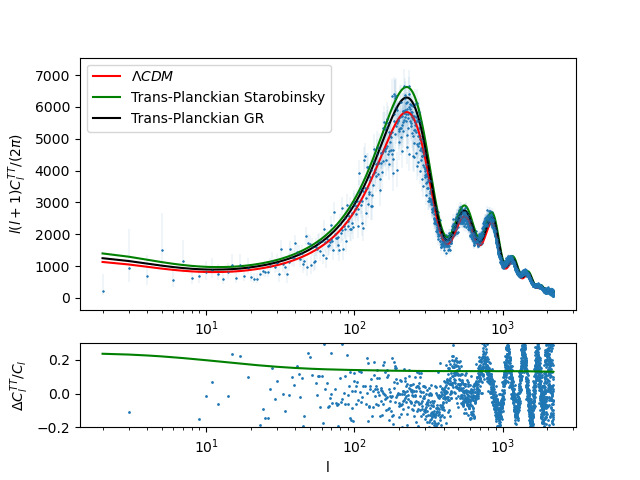}
 \caption{\footnotesize{CMB power spectra TT mode for the $\Lambda$CDM ,trans-Planckian GR and trans-Planckian Starobinsky  models using the best fit parameters to the Planck 2018 data. The green line is based on the best fit parameters of Table 1. The black line is based on the best fit parameters of reference \cite{Bouzari nezhad}. The blue dots are observational data from the Planck 2018 baseline high-$l$ power spectra. The error bars show the $\pm1\sigma$ diagonal uncertainties. In the lower panel, the residual of the trans-Planckian Starobinsky model and the observational data are plotted with respect to $\Lambda CDM$.}}
  \label{fig:TT}
  \end{figure}

\begin{figure}
  \centering
  \includegraphics[width=12cm]{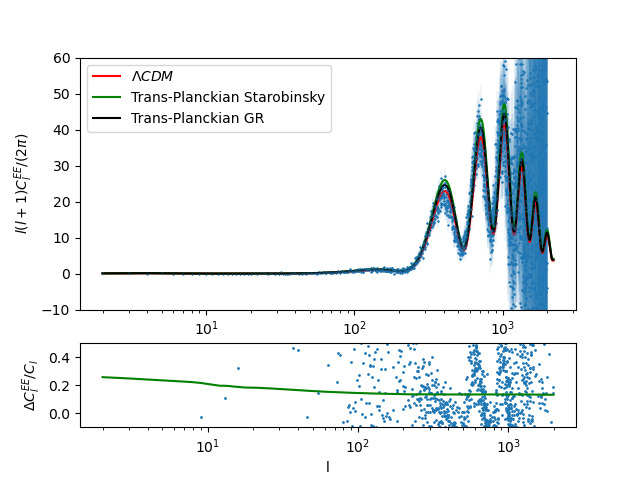}
  \caption{\footnotesize{CMB power spectra EE mode. The red line shows the $\Lambda$CDM with parameters of Planck 2018, the green line is based on the parameters of Table 1 for the trans-Planckian Starobinsky and the black line shows the modes for the trans-Planckian GR using the parameters derived in reference \cite{Bouzari nezhad}. The blue dots are observational data from the Planck 2018 baseline high-$l$ power spectra. The error bars show the $\pm1\sigma$ diagonal uncertainties. In the lower panel, the residual of the trans-Planckian Starobinsky model and the observational data are plotted with respect to $\Lambda CDM$.}}
  \label{fig:EE}
  \end{figure}

\begin{figure}
  \centering
  \includegraphics[width=12cm]{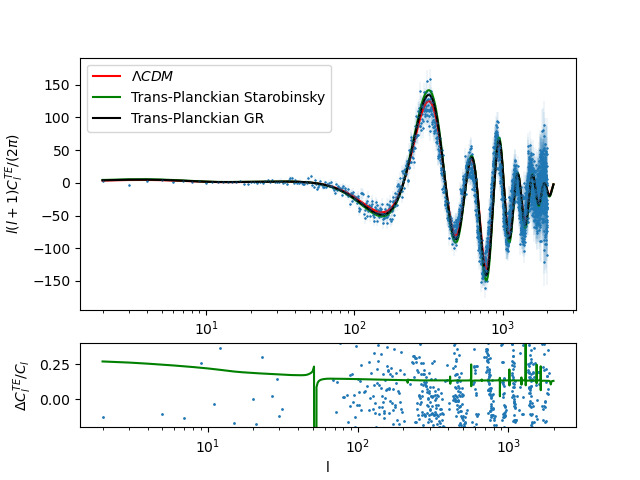}
  \caption{\footnotesize{CMB power spectra TE mode for the $\Lambda$CDM ,trans-Planckian GR and trans-Planckian Starobinsky  models using the best fit parameters to the Planck 2018 data.  The blue dots are observational data from the Planck 2018 baseline high-$l$ power spectra. The error bars show the $\pm1\sigma$ diagonal uncertainties. In the lower panel, the residual of the trans-Planckian Starobinsky model and the observational data are plotted with respect to $\Lambda CDM$.}}
  \label{fig:TE}
\end{figure}

\begin{figure}
  \centering
  \includegraphics[width=12cm]{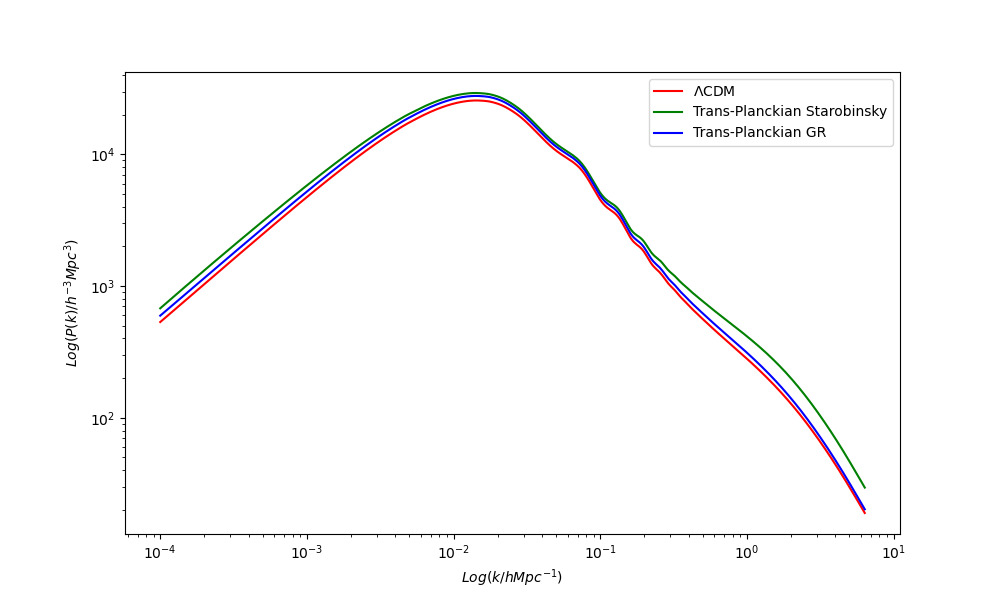}
  \caption{\footnotesize{CMB matter power spectrum for the $\Lambda$CDM (red) using the parameters of Planck 2018, the trans-Planckian GR model (blue) with the best-fitting parameters from \cite{Bouzari nezhad} 
  and trans-Planckian Starobinsky model (green) using best fit parameters of table 1}}
  \label{fig:Matter}
\end{figure}

In this section, we compare the results with the observational data. For this purpose, we have used two commonly used codes. The first is the MGCAMB (Modification of Growth with CAMB) code which is a modified CAMB code. It solves the Boltzmann hierarchy equations  and derives acoustic peaks in modified gravity theories which are parameterized in the form of MoG (Modification of Growth) models \cite{Hojjati},\cite{Zucca},\cite{Antony Lewis}.
The second is the MGCosmoMC code which is also a modification of the CosmoMC code to include MoG models \cite{Antony Lewis Sarah Bridle},\cite{Zhao:2008bn} and is used to fit the parameters of our model to the Planck 2018 results \cite{Planck:2018vyg}.
To obtain the best fit values of the parameters, we rewrite the Starobinsky's power spectrum, equation (\ref{PS}), in the following form
\begin{align}\label{pri}
\begin{split}
&\mathcal{P}^\alpha_\mathcal{R}(k)\simeq \mathcal{A}^{\alpha}_s\big(\frac{k}{k_l}\big)^{n^{\alpha}_s-1}\bigg[1+0.92\epsilon_{1f}+2.92\eta_f-2\epsilon_1\ln{\frac{k}{k_l}}\cos\big(2\gamma_0^{-1}(\frac{k}{k_f})^{\epsilon_1})\big)+\gamma_0\sin\big(2\gamma_0^{-1}(\frac{k}{k_f})^{\epsilon_1}\big)\bigg]
\end{split}
\end{align}
where $\mathcal{A}^{\alpha}_{s}=\kappa^2/(576\mu_0\pi^2\epsilon_{1l}^2)$ and $\gamma_0=H_f/\Lambda$. From (\ref{epsilonl epsilonf}), we can see that  $\epsilon_{1l}\simeq\epsilon_{1f}$ up to the leading order in $\epsilon_1$ and $\eta$. We have ignored the terms with the second order slow roll parameters in the spectral index. Also, by noting that $\gamma_0\ll1$, we have taken into account the leading order terms of this coefficient. The pivot scales are set to: $k_f=0.01$ ${Mpc}^{-1}$ and $k_l=0.05$ ${Mpc}^{-1}$. 
It is noteworthy that the fluctuating behavior of the primary power spectrum is compared with the other models with a similar shape of the power spectrum function but with a different origin.
In the literature there are extensive studies of models with different types of features and their observable signatures on CMB power spectra. The authors of \cite{Braglia:2022ftm} categorized feature models (the models with deviations from the slow-roll inflationary ones) into two general families: sharp feature which is corresponding to an instantaneous deviation from the slow roll evolution and resonant feature with a periodic deviation from the slow roll. In both types of features, the primordial power spectrum of the scalar fluctuations has the oscillating behavior. These features can be described by the form of the Lagrangian (see \cite{Braglia:2022ftm} for more details). As some examples, we can mention: Classical Primordial Standard Clocks (CPSC) \cite{CPSC}, resonance model with a periodic modulation to the slow-roll potential, bump or dip model in which the potential of the inflaton field is a Gaussian profile and , turn model \cite{turn}.
Comparing figure \ref{fig:prim ps} with figure 4 in reference \cite{Braglia:2022ftm}, we see that the behavior of the primordial power spectrum in our model is very similar to the bump models but in our case the wavelength of the oscillations is much longer. This can be justified by comparing equation (B.18) in \cite{Braglia:2022ftm} with our derived spectrum of \eqref{pri}. It seems reasonable because it is known that the Lagrangian of the Starobinsky model can be reparametrized into the  exponential form of the potential in the Jordan frame.\\
For our derived power spectrum, we run MGCosmoMC to find the best fitting values of the parameters for our model. These are shown in Table 1.
\begin{table}
\begin{center}
	\begin{tabular} { |c|c|c|c|c|c| }
	\hline
	Parameter  & best fit & mean & $\sigma$ & $99\% $ lower & $99\%$ upper \\
	\hline
	$\log A_s^{\alpha}$ &3.1066 & 3.1413 & 0.1251 & 2.8656 & 3.5581   \\
	\hline
	$n_s^{\alpha}$ & 0.9612& 0.9514& 0.0185 &  0.9321 & 0.9821\\
	\hline
	$\epsilon_1$& 0.0111 & 0.0102  &0.0074 & 0.0013 & 0.02932\\
	\hline
	$\eta_f$& 0.0200 & 0.01765  & 0.0106 & 0.0015  & 0.05432 \\
	\hline
	$\gamma_0$& 0.04324 & 0.03577&  0.010 & 0.0056 & 0.05123 \\
	\hline
	$\epsilon_{1f}$ & 0.0104 & 0.0132 & 0.00906  & 0.0012 & 0.0400  \\
	\hline
	\end{tabular}
	  \caption{Best-fit values of the parameters of the trans-Planckian Starobinsky model using MGCosmoMC}
\end{center}
\end{table}
The primordial power spectrum may leave detectable traces in the late-time observables, CMB and matter fluctuations. Given the best-fit values of the parameters, we can use the MGCAMB program to plot the primordial power spectrum,  $\mathcal{P}_{\mathcal{R}}^\alpha(k)$ and also the multipole coefficient of $C_{TT}$, $C_{EE}$ and $C_{TE}$ which are respectively temperature-temperature (TT), polarization-polarization (EE) correlation functions, and the temperature polarization (TE) cross-correlation function, in terms of the multipole moment $l$. The result is the famous CMB acoustic peaks. 
We also compare the trans-Planckian Starobinsky model with the trans-Planckian GR and $\Lambda$CDM in our plots. In \cite{Bouzari nezhad}, the authors have studied the effects of $\alpha$-vacua initial states on the power spectrum of the scalar and tensor perturbations during inflation in GR. They have found the primordial power spectrum with the best fitting values of the free parameters (in the case of GR, the slow roll parameters and $\gamma_0$ are the free parameters).
In addition, we present the matter power spectrum for the three models mentioned above.\\
Now, let us have a brief review of the parameterization scheme in the MGCAMB code and introduce some functions to modify gravity and the contribution of an exotic dark energy fluid \cite{Zhao:2008bn}.
Note that in $f(R)$ gravity, both the background and the perturbation equations are affected by the model under consideration, so Einstein’s equations and the relation between the scalar degrees of freedom of the metric perturbations should also be modified. Imposing the Newtonian gauge condition, we follow the general parametrization of the scalar metric perturbations, i.e. the gravitational potentials proposed in \cite{Zhao:2008bn} to show deviations from GR.  In modified gravity, these potentials can have scale-dependent growth patterns and they are not necessarily equal. Following \cite{Zhao:2008bn}, we have parameterized the ratio of the gravitational potentials and the Poisson equation of $\Psi$ as follows\footnote{This parameterization follows from the scalar-tensor theories and also, from the relations between the scalar perturbations of the metric in Einstein and Jordan frames. It is also assumed that this form of the parameterization functions holds only in the linear regime of the cosmological density fluctuation, and that $\mu$ and $\gamma$ tend to unity on solar system scales.}
\begin{eqnarray}\label{f}
k^2\Phi=-\frac{1}{2} a^2\mu(k,a)\rho\Delta \hspace{0.5in} \frac{\Psi}{\Phi}=\gamma(k,a)
\end{eqnarray} 
where $\Phi$ is the temporal scalar perturbation of the metric, $\mu(k,a)$ and $\gamma(k,a)$ are two scale- and time-dependent functions that encode any late time modification of the Poisson and the anisotropic stress equations of GR. The comoving density perturbation $\Delta$ satisfies 
\begin{equation}
\rho\Delta=\rho\delta+3\frac{\mathcal{H}}{k}(\rho+P)v
\end{equation}
where $\delta=\delta\rho/\rho$ is the density contrast.
By virtue of (\ref{f}),  the results are expected to depend on the choice of the parametrization functions  $\mu(k,a)$ and $\gamma(k,a)$. However, one can use a special parametrization that accurately represents a broad class of modified gravitational theories \cite{Zhao:2008bn}, \cite{Edmund Bertschinger}: 
\begin{equation}
\mu=\frac{1+\beta_1\lambda_1^2k^2a^s}{1+\lambda_1^2k^2a^s}\hspace{0.5in}\gamma=\frac{1+\beta_2\lambda_2^2k^2a^s}{1+\lambda_2^2k^2a^s}
\end{equation}
where for the $f(R)$ theories, it can be shown that $\beta_1=4/3$, $\beta_2=0.5$, $\lambda_2^2=\beta_1\lambda_1^2$.  According to \cite{Zhao:2008bn}, $s=4$ for a viable $f(R)$ model. Thus, the only free parameter is $\lambda_1$ and it would be related to the mass scale of the scalar degree of freedom introduced by the $f(R)$ function. This parameter is constrained from below by the requirement of consistency with the cosmological and local tests.
The valid bound from the cosmological tests is $\lambda_1^2\leq 10^6Mpc^2$ and from the local tests is $\lambda_1^2\leq10^2 Mpc^2$. We set $\lambda_1^2\sim 10^3 Mpc^2$ for the Starobinsky model. We have plotted the primordial power spectrum for the $\Lambda$CDM (red), trans-Planckian GR (green) and trans-Planckian Starobinsky (blue) models in Figure (\ref{fig:prim ps}). This figure shows that for the smaller value of the wavenumber, the trans-Planckian Starobinsky plot intersects the $\Lambda$CDM and trans-Planckian GR plots. The oscillation rate of the primordial power spectrum in the trans-Planckian Starobinsky model is milder than in the trans-Planckian GR and there is a small deviation between these models for large values of the wavenumber.\\
In Figures (\ref{fig:TT}-\ref{fig:TE}), using the different types of angular power spectra including the TT, EE and TE correlation functions, we have studied the temperature and polarization anisotropy of the CMB for three models: trans-Planckian GR (blue), trans-Planckian Starobinsky (green) and $\Lambda$CDM (black). Blue dots in the plots indicate the observational data from the Planck 2018 dataset of baseline high-$l$ power spectra. The error bars are $\pm1\sigma$ uncertainties in the main plots. The horizontal axis is the angular parameter $l$, on a logarithmic scale. The lower panel shows the residuals. These are the relative difference of the trans-Planckian Starobinsky model with respect to $\Lambda$CDM ($($trans-Planckian Straobinsky - $\Lambda$CDM$)$/$\Lambda$CDM). In addition, the blue points in the lower panels are the residuals of the observed points. To improve the clarity of the plots, we omit the error bars in the lower panels.
Our results can be compared to the case of the standard $\Lambda$CDM model. It is clear from these plots that there is a significant difference between the multipole coefficients in the amplitude of the oscillatory peaks.
It is also worth noting that our results converge to those obtained for the $\Lambda$CDM model for small and large values of $l$. However, there is some deviation from the $\Lambda$CDM model in the range of intermediate values of $l$.
We have plotted the matter power spectrum, $P(k)\sim\left\langle\delta_m^2\right\rangle$, at $z=0$ for $\Lambda$CDM, trans-Planckian GR and trans-Planckian Starobinsky models in Figure (\ref{fig:Matter}) by using the best fit values of the parameters in Table (1) and also the parameters mentioned in MGCAMB.  
We see that the value of the matter power spectrum in both trans-Planckian GR and trans-Planckian Starobinsky models, is more than what results from the $\Lambda$CDM model.  Moreover, for sufficiently large values of the wavenumber, the difference between these two trans-Planckian models increases.
The power spectrum of matter in the trans-Planckian GR is smaller than that of $\Lambda$CDM and larger than that of the trans-Planckian Starobinsky at all scales while the relative differences between these models vary with scale. 
\pagebreak
\section{Conclusion}
The initial state of the quantum fluctuations can lead to slightly different predictions for the late time cosmological observables such as the CMB multipoles and the matter power spectrum. 
In the absence of a satisfactory quantum gravity, we can parameterize the trans-Planckian effect with some modified vacuum states. The various vacuum choices include the $\alpha$-vacuum \cite{Danielsson:2002kx},
the coherent state \cite{S Kundu1}\cite{S Kundu2}, the $\alpha$-states \cite{E Mottola}\cite{B Allen}, the thermal state \cite{K Bhattacharya}, and the excited-de Sitter modes \cite{E Yuso}.  It is interesting to investigate whether there is a feature of trans-Planckian physics that can be observed in the current cosmological data.\\
Here we have used the $\alpha$-vacuum as an alternative to the usual BD vacuum. It has been shown that the Hubble parameter and also the initial time are $k$-dependent quantities. We have introduced two pivot scales: $k_l$ is the last wavenumber leaving the horizon and $k_f$ is the first wavenumber satisfying the initial condition.
Using these two scales and also, imposing an ultraviolet cutoff $\Lambda$ on the physical momentum, we have obtained the slow roll and Hubble parameters at the initial and horizon-crossing times. Taking into account $f(R)$ gravity, we have calculated the corrections to the scalar power spectra and compared the results with those obtained by assuming the initial BD vacuum. \\
The resulting power spectrum contains an oscillatory term 
with a $k$-dependent amplitude proportional to $H_f/\Lambda$. This term is called the trans-Planckian correction.
Our results are in agreement with the previous works \cite{Lim}, \cite{Bouzari nezhad} , which considered the non-BD initial conditions in GR. We have shown the best fitting values of the parameters of our models
and also, made comparisons with the $\Lambda$CDM model in Figures (\ref{fig:prim ps}-\ref{fig:Matter}). We found  amplitude enhancement in the CMB, TT, TE and EE modes. Also, the matter power spectrum in the trans-Planckian Starobinsky is always larger than the corresponding one in the GR. This result is  in agreement with \cite{Chen:2019uci} where two viable models of $f(R)$ gravity, exponential and Starobinsky models, are considered and the authors show that the enhancement of the matter power spectrum is a generic property of these models. Our results show that the matter power spectrum in the Starobinsky model with the BD initial state  is larger than the corresponding one in the GR. \\
For the future directions, it would be interesting to consider other initial vacuum states and obtain their corrections to the scalar power spectrum. Furthermore, the corrections of different initial states of inflationary fluctuations can be studied at higher moments, such as the bispectrum.
\subsection*{Acknowledgements}
We are  very  grateful to A. Jozani and H. Bouzari Nezhad for their fruitful collaboration on this paper. S. Ch. is supported by the Transilvania Fellowship Program, 2022. F. S. is grateful to the University of Tehran for supporting this work with a grant from the University Research Council.

\end{document}